\documentclass[pdflatex,sn-mathphys-ay]{sn-jnl}

\usepackage{graphicx}%
\usepackage{multirow}%
\usepackage{amsmath,amssymb,amsfonts}%
\usepackage{amsthm}%
\usepackage{mathrsfs}%
\usepackage[title]{appendix}%
\usepackage[dvipsnames]{xcolor}%
\usepackage{textcomp}%
\usepackage{manyfoot}%
\usepackage{booktabs}%
\usepackage{algorithm}%
\usepackage{algorithmicx}%
\usepackage{algpseudocode}%
\usepackage{listings}%
\usepackage{aas_macros}
\usepackage{wasysym}
\usepackage{mathrsfs}

\usepackage{scalerel}
\usepackage{tikz}
\usetikzlibrary{svg.path}

\definecolor{orcidlogocol}{HTML}{A6CE39}
\tikzset{
  orcidlogo/.pic={
    \fill[orcidlogocol] svg{M256,128c0,70.7-57.3,128-128,128C57.3,256,0,198.7,0,128C0,57.3,57.3,0,128,0C198.7,0,256,57.3,256,128z};
    \fill[white] svg{M86.3,186.2H70.9V79.1h15.4v48.4V186.2z}
                 svg{M108.9,79.1h41.6c39.6,0,57,28.3,57,53.6c0,27.5-21.5,53.6-56.8,53.6h-41.8V79.1z M124.3,172.4h24.5c34.9,0,42.9-26.5,42.9-39.7c0-21.5-13.7-39.7-43.7-39.7h-23.7V172.4z}
                 svg{M88.7,56.8c0,5.5-4.5,10.1-10.1,10.1c-5.6,0-10.1-4.6-10.1-10.1c0-5.6,4.5-10.1,10.1-10.1C84.2,46.7,88.7,51.3,88.7,56.8z};
  }
}

\newcommand\orcidicon[1]{\href{https://orcid.org/#1}{\mbox{\scalerel*{
\begin{tikzpicture}[yscale=-1,transform shape]
\pic{orcidlogo};
\end{tikzpicture}
}{|}}}}

\theoremstyle{thmstyleone}%
\theoremstyle{thmstyletwo}%

\theoremstyle{thmstylethree}%

\begin{document}

\title[Article Title]{A 5kHz Modulator for Pyramid Wavefront Sensors}


\author*[1,2]{\fnm{Maximilian} \sur{Häberle \orcidicon{0000-0002-5844-4443}     }}\email{haeberle@mpia.de}

\author[1]{\fnm{Vianak} \sur{Naranjo}}
\author[1]{\fnm{Yared} \sur{Reinarz}\orcidicon{0000-0003-1932-3494}}

\author[1]{\fnm{Markus} \sur{Feldt}\orcidicon{0000-0002-4188-5242}}

\author[1]{\fnm{Silvia} \sur{Scheithauer}\orcidicon{0000-0003-4559-0721}}

\author[1]{\fnm{Thomas} \sur{Bertram\orcidicon{0000-0003-0548-9709}}}

\author[3]{\fnm{Arne} \sur{Bramigk}}

\author[3]{\fnm{Harry} \sur{Marth}}

\affil[1]{\orgname{Max Planck Institute for Astronomy (MPIA)}, \orgaddress{\street{Königstuhl 17}, \postcode{69117 }\city{Heidelberg}, \country{Germany}}}

\affil[2]{\orgname{European Southern Observatory}, \orgaddress{\street{Karl-Schwarzschild-Straße 2}, \postcode{85748 }\city{Garching},  \country{Germany}}}

\affil[3]{\orgname{Physik Instrumente (PI) SE \& Co. KG}, \orgaddress{\street{Auf der Römerstraße 1}, \postcode{76228 }\city{Karlsruhe}, \country{Germany}}}


\abstract{Despite the emergence of new types of wavefront sensors, the modulated pyramid wavefront sensor remains the workhorse for ELT instrumentation, and is among the options even for advanced high-contrast, high-Strehl instrumentation like PCS and SAXO+.  To achieve the required degree of wavefront control, an operation at frequencies of 3kHz, ideally up to 5kHz, is necessary, requiring an optomechanical device capable of delivering accurate circular modulation patterns with these frequencies. 
Here, we present tests of a novel type of high-frequency modulator based on shearing piezo actuators.
The modulator prototype moves a flat circular mirror (15mm diameter) with a tip-tilt range of plus/minus 50 arcsec.  At a typical 10mm pupil diameter on the modulator mirror, and operating at 2.2\,µm, this will create a modulation circle with a radius of slightly greater than 2$\lambda$/D.  While this is less than conventionally specified for most instruments, it should already be sufficient for any practical application except for very bad conditions or extended targets.
We performed modulation tests at frequencies between 250 Hz to 5 kHz using a test setup including a modulated laser beam probed with a high-speed camera. The prototype showed stable behaviour during a one-hour-long operation at a maximum frequency of 5 kHz and with negligible heat generation. The maximum modulation amplitude was 60\,arcsec. We observed very accurate reproduction of the input modulation pattern with typical ellipticities less than 1\% and random deviations below 0.2\% for frequencies below 4.5\,kHz.
These tests demonstrate the prototype's capabilities and could be followed by on-sky tests or the integration of the modulator into XAO testbeds.}


\maketitle

\section{Introduction}\label{sec1}

Adaptive optics (AO) have enabled diffraction-limited observations with ground-based telescopes, leading to various breakthroughs in our astrophysical understanding of the universe \citep{2012ARA&A..50..305D}, from high-resolution studies of the Galactic Center \citep[e.g.][]{2017ApJ...837...30G} to direct observations of exoplanets \citep[e.g.][]{sphere-shine}. With future instruments and even larger telescopes, the requirements on AO systems are continuously evolving. In particular the direct observation of exoplanets, poses the most demanding requirements on AO systems.  The need to suppress the light of a planet's host star to detect the planetary source, $10^5$ to $10^9$ times fainter, at separations only a couple of times larger than the theoretical diffraction limit of the telescopes, requires the wavefront entering the instrument to be controlled - and thus determined - down to levels of a few tens of nanometers.  To achieve such a degree of control, measuring and correcting at dense spatial and temporal sampling is required.

One key component of every AO system is its wavefront sensor (WFS), which measures the shape of the incoming wavefront and provides the feedback for the closed-loop control systems. Many current and near-future AO systems use so-called Pyramid Wavefront Sensors (PWFSs) \citep{1996JMOp...43..289R}. While alternative sensor types, such as the Bi-O-edge \citep{2024A&A...682A..27V}, the Zernike-sensor \citep{2024A&A...692A.157N}, or the more elaborate concepts for coronagraphic wavefront sensors \citep{Chambouleyron:24} are emerging and slowly making there way from theory via lab experiments to on-sky operation, the PWFS remains the workhorse for the near-future generation of ELT instruments \citep{2024ExA....58...20F,2022SPIE12185E..4SC,10.1117/12.3019962}, and an important fallback solution for the next generation.

\subsection{Pyramid wavefront sensors \label{sec:pyramid-wfs} and the need for modulation}

The concept of the PWFS was introduced by \cite{1996JMOp...43..289R}.  At the time, the standard WFS in use for astronomical AO systems was the Shack-Hartmann wavefront sensor (SHWFS) \citep{hartmann1900}. The PWFS has a higher sensitivity than the SHWHS, in particular in the low flux regime, and at low spatial frequencies \citep{10.1111/j.1745-3933.2005.08638.x}.  The downside is a limited linear range when compared to the SHWFS.  This downside can be overcome by means of a mechanical modulation device in the form of a tip-tilt mirror at the expense of some of the sensitivity gains \citep{Burvall:06}.  The task of the tip-tilt mirror is to move the PSF image, which in the original concept falls onto the vertex of the four-sided pyramid prism, around said vertex on a full circle once during each integration cycle.  By adjusting the radius of the circle, the so-called modulation amplitude, one can adjust the sensitivity of the sensor versus the linear range: High modulation amplitudes essentially make the PWFS equivalent to the SHWFS, low amplitudes or no modulation at all render the PWFS a very sensitive phase sensor with a very limited dynamic range.  For so-called eXtreme AO (XAO) applications in the context of exoplanet imaging, it is usually found that a modulation amplitude between 2 and 4 $\lambda/D$ is ideal, $\lambda$ being the wavelength at which the sensor is operating, and $D$ the diameter of the telescope's primary mirror 

One current example for such an XAO instrument at the forefront of technological devolvement is the SAXO+ upgrade \cite{2022SPIE12184E..1SB} for the SPHERE high-contrast instrument. It will equip the instrument with a fast second stage AO that will use a PWFS and will run at frequencies of up to 3\,kHz with an ideal modulation amplitude of 2$\lambda/D$ \citep[see][ for detailed performance simulations]{2024A&A...689A.199G}. SAXO+ is also regarded as a mandatory technology demonstrator for the future Planetary Camera and Spectrograph (PCS) at the European Southern Observatory's Extremely Large Telescope. To achieve the necessary wavefront quality, PCS will also use a multistage AO with an even higher sampling rate of  up to 4\,kHz \citep{2021Msngr.182...38K}. While SAXO+ is already in the stage of manufacturing and system assembly, the currently foreseen modulation solution for the PWFS, based on a Physik Instrument (PI) S331 actuator \citep[e.g.]{2022JATIS...8d9001S} in combination with a custom controller, still needs to be verified at high frequencies, and the final achievable frequency is not yet known. A discussion of the requirements for the SAXO+ modulator in context of the tested prototype is given in \autoref{sec:summaryandconclusions}.

PCS on the other hand is in a very early stage, where not even full end-to-end simulations yet exist, and a choice of WFS type cannot be made. Nevertheless, the evidence of the significant challenge frequencies above 1.5\,kHz, let alone the aforementioned 4\,kHz pose, led to a preliminary exclusion of the PWFS type, which can now be reconsidered.

Apart from alternative sensor concepts, modulation-less solutions are also under investigation for PWFSs, see e.g. \cite{2002OptCo.208...51R}, \cite{2003SPIE.4839..288C}, \cite{Hutterer:18}, \cite{Guzman:24}, and \cite{2025A&A...696L...1L}.
Nevertheless, having a modulation device capable of delivering reliable modulation at up to 4\,kHz remains highly desirable, as it would allow the continued operation of the well-proven PWFS concept, of which various instances are in use every night at observatories around the globe, in future planet-hunting XAO systems.

Laboratory experiments that test very fast control loops are e.g. the LLAMAS experiment \citep{2018SPIE10703E..1NA} or the GHOST testbench \citep{10.1117/12.2630595}.

\subsection{Outline}
In this paper, we describe tests performed on a new high-frequency modulation device capable of modulation frequencies of 5\,kHz. In \autoref{sec:device} we describe the prototype device itself. In \autoref{sec:setup} we describe the optical setup used to test the new modulator. \autoref{sec:datareduction} describes the data reduction and the results of the modulation performance tests. We conclude the paper with a summary (\autoref{sec:summaryandconclusions}) and possible next steps in \autoref{sec:future}.

\section{Description of the modulation device}
\label{sec:device}
\begin{figure}
    \centering
    \includegraphics[width=0.35\linewidth]{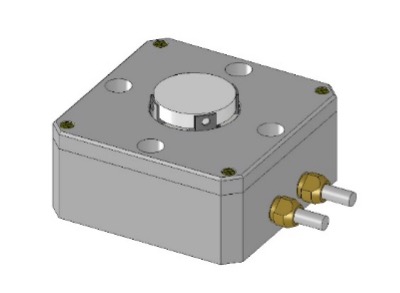}
    \includegraphics[width=0.35\linewidth]{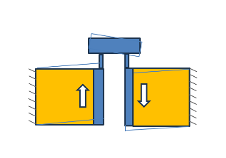}
    \caption{\textit{Left:} 3D Design drawing of the high-frequency modulator prototype. \textit{Right: } 2D Schematic showing the push-pull actuation principle.}
    \label{fig:modulator_design}
\end{figure}

The high-frequency modulator prototype was designed and built by Physik Instrumente (PI) in Karlsruhe. The basic specifications are given in \autoref{tab:specification}. 
A design with four piezo actuators was chosen to achieve the required high dynamics. These are arranged in a $4\times90^\circ$. circle, with the opposite actuators being controlled in push-pull mode (see \autoref{fig:modulator_design}). This results in a counter-rotating 
movement of the actuators and the desired tilting movement of the mirror platform in two orthogonal axes.

\begin{table}
    \centering
        \caption{Specifications of the modulator prototype}
    \label{tab:specification}
    \begin{tabular}{|ll|}
    \hline
        External dimensions:\hspace{0.4cm} & $40\times40\times18$\,mm$^3$\\  
        Mirror dimensions:  & $\diameter$15\,mm$\times$3\,mm\\
        Max. Tip-Tilt angles: & $\pm34$\,arcsec at $\pm500$\,V\\
        Resonance frequency: & 9.8\,kHz\\
    \hline
    \end{tabular}

\end{table}

The piezo actuators are manufactured in a stacked design from PIC050 piezoelectric material (PI Ceramic). This crystalline material exhibits excellent linearity of deflection with negligible hysteresis, meaning that in many cases a position sensor is not required and open-loop operation is possible. A shear actuator was chosen as the design. The electro-mechanical coupling of the piezoelectric shearing mode is highly efficient, enabling a compact design that meets the desired stroke, with high stiffness to promote good dynamical properties.

Piezo actuators made of PIC050 behave electrically like capacitors with very low capacitance and low dielectric loss factors. The Reactive power does not heat up the actuators and  allows continuous operation at high frequencies. The above application has been tested for continuous operation up to 5\,kHz without significant heating of the mechanics. 

\section{Measurement setup}\label{sec2}
\label{sec:setup}
\subsection{Optical setup}
\begin{figure}
    \centering
    \includegraphics[width=1\linewidth]{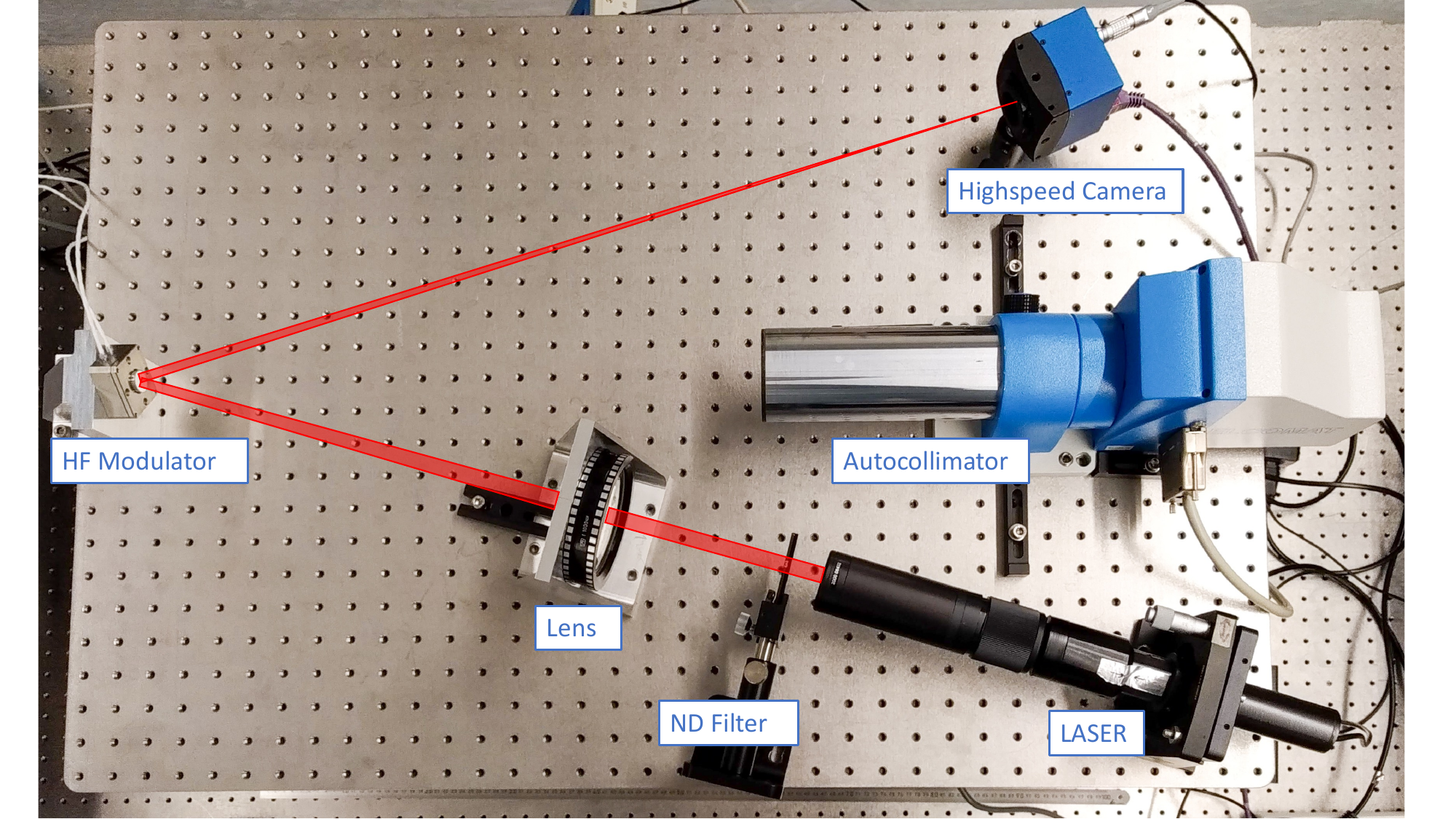}
    \caption{Overview of the test setup mounted on an optical table. All components are labeled. The path of the laser beam is indicated in red.}
    \label{fig:overview}
\end{figure}

To verify the performance of the modulator at high frequencies at MPIA, we used a test setup in which we tracked the path of a LASER beam deflected by the HF modulator using a highspeed camera. An overview of the different components that were all mounted on an optical table is given in \autoref{fig:overview}. A description of the different optical components (in the order they appear in the light path) is given in the following.
\begin{enumerate}
    \item A 1mW HeNE LASER (Manufacturer: Melles-Griot) with a $\times30$ beam expander is used as light source and produces a collimated beam of width of approximately 30\,mm.
    \item The light is attenuated by 99.99\% using a combination of neutral-density (ND) filters to avoid saturating the camera.
    \item A lens with a focal length of $f=1000\,$mm is used to focus the beam on the detector of the camera
    \item The converging light beam falls on the modulator, which has a movable, circular mirror with a diameter of 15mm. The light is reflected with the commanded tip-tilt offset.
    \item The focused light finally falls on the sensor of a high-speed camera (\textit{Mikroton MotionBLITZ EoSens mini2}) which is continuously read out at high frequency to record the modulation patterns.
\end{enumerate}

\subsection{Measurement process to characterize the modulation pattern}
To command the modulator, a 2-channel frequency generator (\textit{KEYSIGHT 33600A Series}) is connected to a high-voltage amplifier\footnote{Amplifier model: \textit{Physik Instrumente: E-508K026}, Voltage amplification: $\times100$}, that produces the high-voltage necessary to drive the two axes of the modulator. In order to produce a circular modulation pattern, we command the modulator with two identical sinusoidal signals shifted by 90$^\circ$. To study the behaviour at different frequencies and amplitudes, we conduct a sweep through 20 different frequencies (from 250\,Hz to 5000\,Hz in steps of 250\,Hz) and 10 different amplitudes (with the amplifier input signal ranging from 1\,V$_{\rm pp}$ to 10\,V$_{\rm pp}$ in steps of 1\,V. The output voltage of the amplifier therefore ranged from 100\,V$_{\rm pp}$ to 1000\,V$_{\rm pp}$.

At each frequency-amplitude combination, we record a series of 1000 images with the highspeed camera. The camera is read out at a frame rate of 36899\,Hz and with a shuttertime of 3\,$\mu$s, meaning that each recorded sequence has a length of around 0.027$\,$s. To achieve the high frequencies, a region of interest of 144$\times$122 pixels is used, enough to capture the full modulated circle even at the highest amplitudes.
\subsection{Calibration of modulator angles using an autocollimator}
\label{subsec:calibration}
To find the precise relation between modulator angles and recorded positions of the modulated LASER beam on the camera sensor, we probe the angle of the modulator mirror at different constant input voltages with an autocollimator (\textit{Möller-Wedel ELCOMAT 3000}), while simultaneously saving the readout of the camera. The autocollimator measurements are performed relative to the resting position of the modulator, i.e. when 0\,V are applied to both axes.

\begin{figure}
    \centering
    \includegraphics[width=1\linewidth]{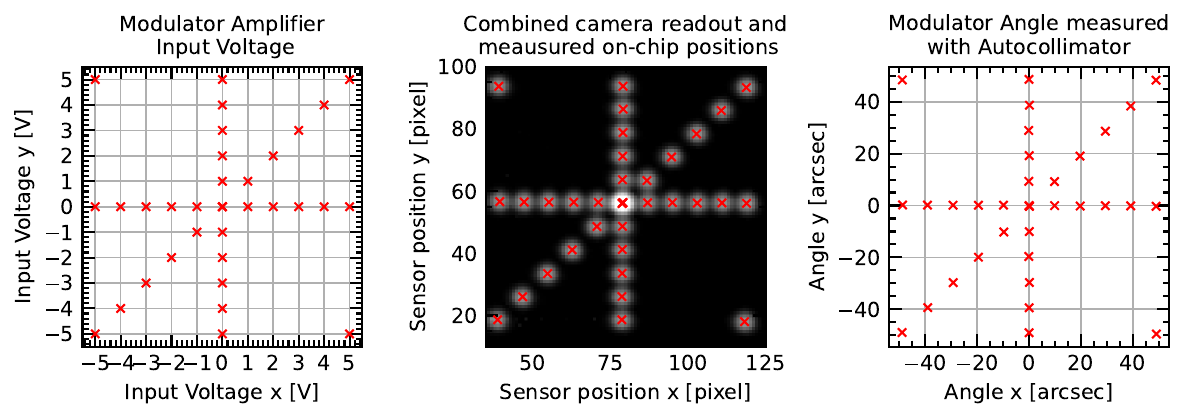}
    \caption{This figure illustrates the three stages of the calibration process. The left panel shows the commanded input voltages, covering the full range used during the modulator operation. The center panel shows the combined readout of the highspeed camera, the laser spot at different positions is clearly visible. Finally the right panel shows the corresponding modulator angles, directly measured with the autocollimator.}
    \label{fig:calibration}
\end{figure}

We used different input voltages covering the full specified range of the modulator (i.e. from $-5$\,V to 5\,V in both axes). An illustration of this process is shown in \autoref{fig:calibration}. Measured angles ranged from $\pm50$\,arcsec, exceeding the specified range (see \autoref{tab:specification}). We then use a least-square fit to determine the optimal affine transformation between pixel coordinates and angles. Using an affine transformation is necessary, as a slight misalignment of the camera sensor can lead to a shearing and therefore apparent ellipticity of the modulation pattern. The affine transformation can be described with a partially populated 3$\times$3 matrix or with 6 explicit parameters. The explicit form as implemented in \texttt{scikit.image} \citep{scikit-image} is:

\begin{align}
    x' &= s_x \cdot x\cdot (\cos{\theta}) - s_y \cdot y\cdot (\sin{\theta}+\tan{h}\cdot\cos{\theta}) + t_x \\
    y' &= s_x \cdot x\cdot (\sin{\theta}) - s_y \cdot y\cdot (\tan{h}\cdot\sin{\theta}-\cos{\theta}) + t_y
\end{align}
with the scale factors $s_x$ and $s_y$, the rotation angle $\theta$, the shear value $h$, and the translations $t_x$ and $t_y$.
The best fit transformation transformation shows minimal rotation ($\theta=$0.12$^\circ$) and a mean scale of $\frac{1}{2}(s_x+s_y)=$1.26 arcsec/pixel. This is matching our expectation: The high-speed camera pixels have a size of 8\,$\mu$m and are at a distance of around 66\,cm from the modulator mirror, leading to an expected scale of 1.25\,arcsec/pixel.
The residuals of our transformations show a root mean square of 0.17\,arcsec, which is within the nominal accuracy of the autocollimator ($\pm$0.25\,arcsec). We note that while the calibration delivers the absolute scale of our measurements, it does not significantly influence the achievable accuracy of the (relative) high-frequency tests.

\subsection{Test of sound emission}
At higher frequencies ($>2000\,$Hz) the modulator produces a noticeable (and eventually disturbing) sound. We measured the acoustic signal with a sound level meter (\textit{VOLTCRAFT Schallpegelmessgerät 320}), mounted approximately 75\,cm above the optical bench. The modulator was run at the highest possible amplitude  (10\,V$_{\rm pp}$) and with frequencies between 250\,Hz and 7000\,Hz. The measured sound levels showed an increase with frequency, and reached a level of around 80\,dB at frequencies $\geq5\,$kHz. We note that the acoustic tests were performed in an optical laboratory without any mitigation of possible acoustic echoes or resonances and are therefore of a qualitative nature.

\subsection{Measurement of long term stability and heat-generation}
\begin{figure}
    \centering
    \includegraphics[width=0.75\linewidth]{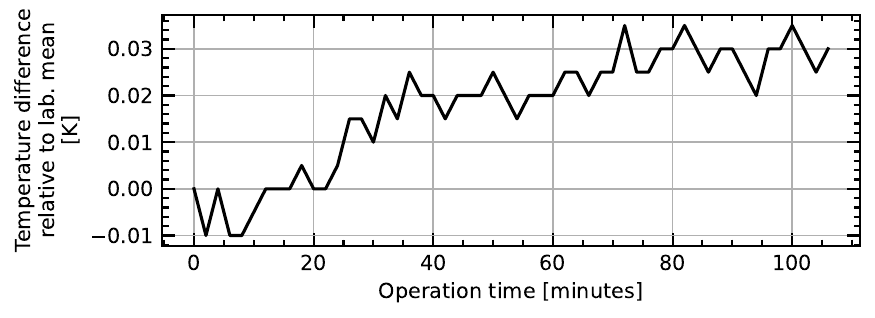}
    \caption{Change of the modulator housing temperature relative to the mean temperature of the laboratory.}
    \label{fig:temperature}
\end{figure}

To test the long-term stability during operation of the modulator, we ran it continuously for 1\,hour with a modulation frequency of 5000\,Hz and at an input amplitude of 10\,V$_{\rm pp}$. At the same time we recorded the temperature of the modulator housing using a DT-670 temperature sensor. \autoref{fig:temperature} shows how the measured temperature of the modulator housing evolved with respect to the mean temperature of two other temperature sensors placed at different locations in the laboratory. Only very minor temperature fluctuations (full range: $\Delta T=0.04$\,K) were observed. The measured fluctuations  were significantly lower than the overall temperature fluctuations within the laboratory. 

\section{Data reduction and results}\label{sec3}
\label{sec:datareduction}
\begin{figure}
    \centering
    \includegraphics[width=1\linewidth]{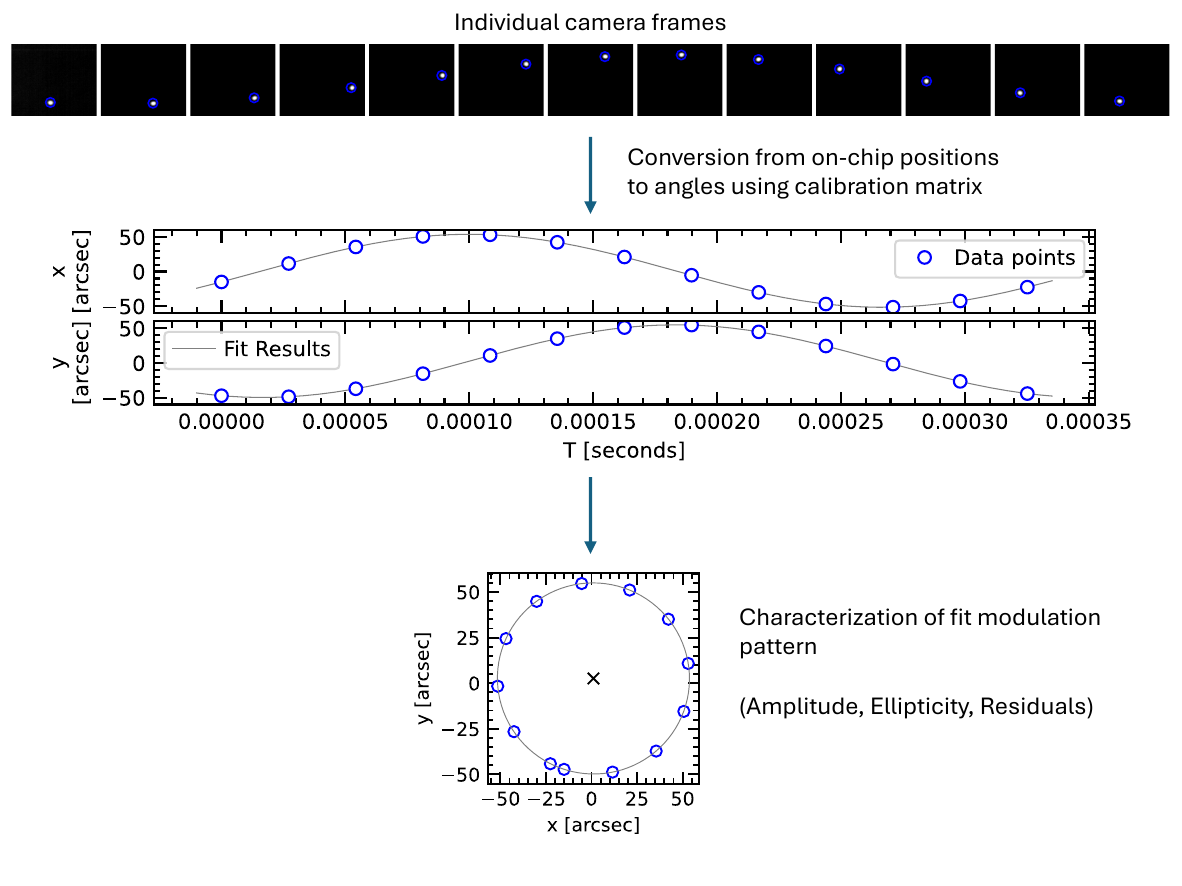}
    \caption{Illustration of the data reduction process that transforms the time series of images measured with the high speed camera to a time-series of 2D angles.}
    \label{fig:data_reduction}
\end{figure}

\subsection{Data reduction}
The goal of this step is to derive the properties of the modulation pattern from the image sequences recorded for different modulator frequencies and amplitudes. An illustration of this process if given in \autoref{fig:data_reduction}.

\subsubsection{Centroiding}
\begin{figure}
    \centering
    \includegraphics[width=0.49\linewidth]{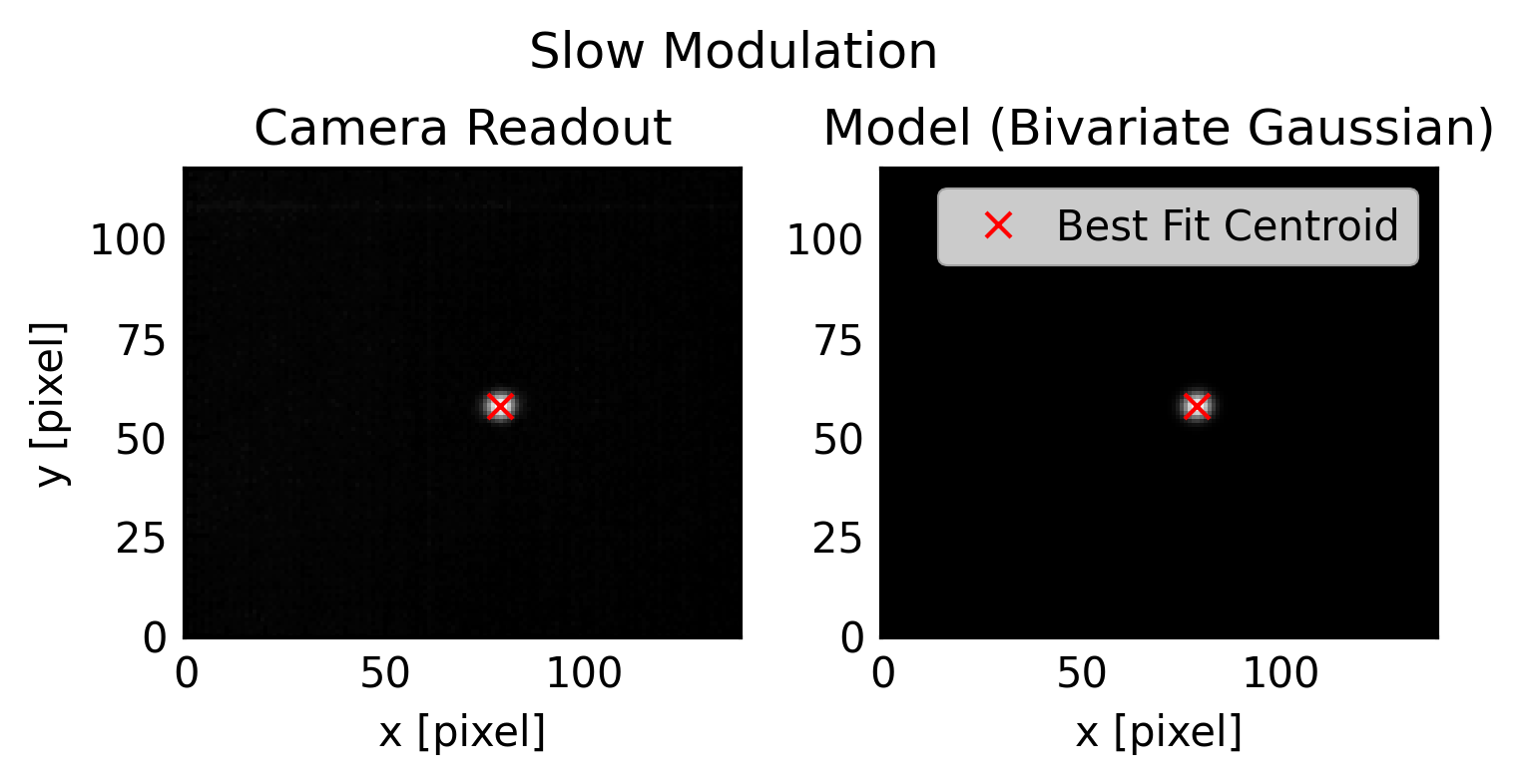}
    \includegraphics[width=0.49\linewidth]{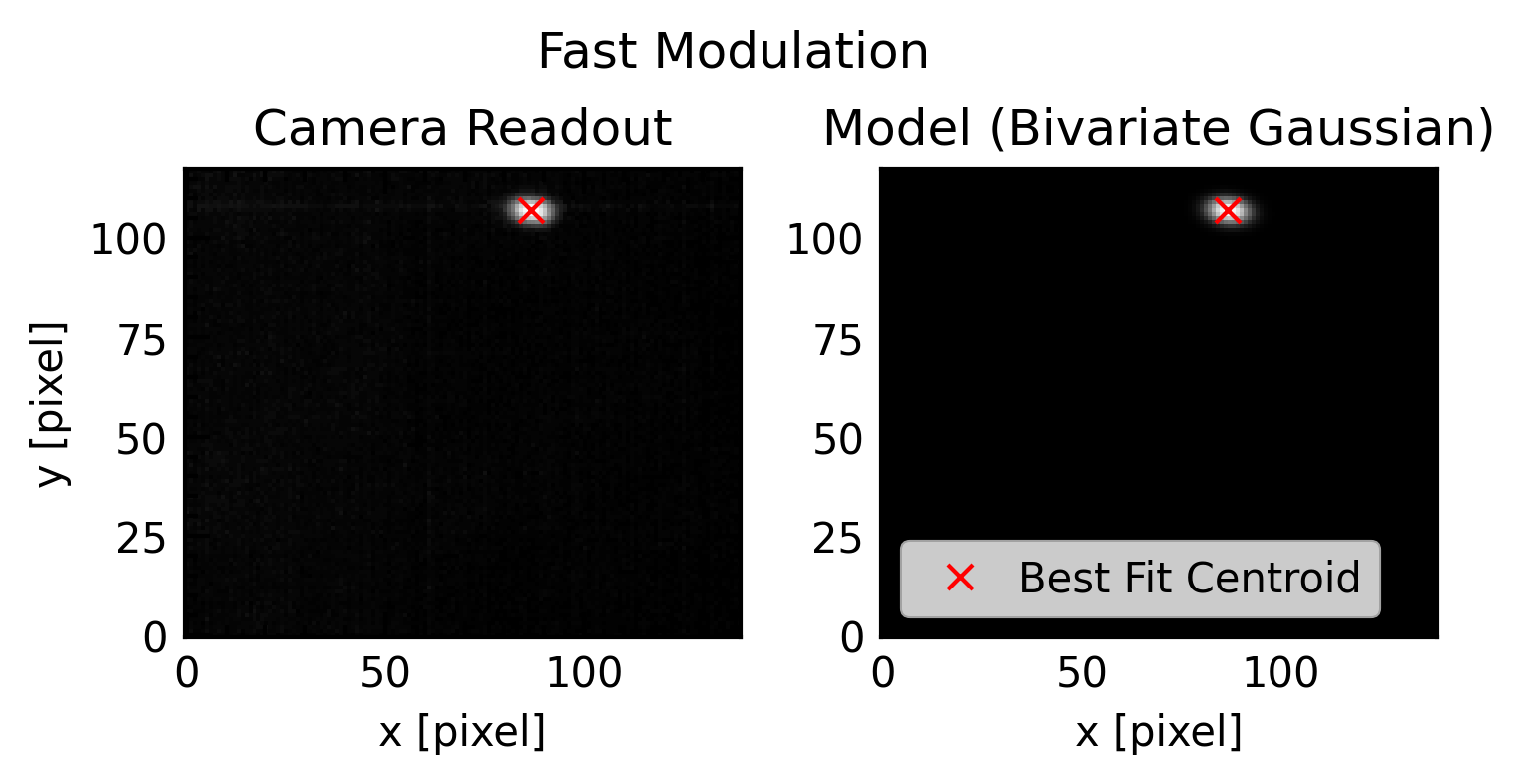}
    \caption{Illustration how centroids are determined in the individual camera frames by fitting a bivariate Gaussian. Here we show two example images and the best fit model recorded during modulation with low frequencies / amplitudes (\textit{left}) and at the maximum frequency/amplitude (\textit{right}). At high frequencies the LASER spot starts to become elongated, but the centroid is still recovered correctly.}
    \label{fig:image_model}
\end{figure}

For each image we start by fitting the LASER spot by fitting a 2-dimensional Gaussian. The central coordinates of the best-fit Gaussian are our best estimate of the on-chip position of the LASER spot. The measured FWHM of the LASER spot is about 5 pixels (4 arcsec). At the highest amplitudes and frequencies it is elongated due to smearing, but even in the most extreme case (Amplitude = 62\,arcsec, $f=$5000\,Hz) the smearing only reaches 6 pixels due to the short exposure time (3\,~µs). The signal-to-noise is high ($\sim300$) and estimating the centroiding precision with $\sigma_\textrm{centroid}\approx\textrm{FWHM / SNR}$ \citep{1978moas.coll..197L} yields an expected precision of $\leq$0.013\,arcsec, similar to what is actually found in the fit residuals at low modulation frequencies (see \autoref{secA1}). In \autoref{fig:image_model} we show two example images at different frequencies and the best-fit Gaussian model.

\subsubsection{Calibration and fit of modulation pattern}
In the next step, we convert the individual on-chip positions to modulator angles using the calibration matrix determined in \autoref{subsec:calibration}. The resulting time-series of 2D angles is then fit with the following parametrization, describing an elliptical modulation pattern with arbitrary rotation:
\begin{equation}
    \begin{pmatrix}
x \\
y 
\end{pmatrix}=
    \begin{pmatrix}
A_x\cdot\cos(2\pi\cdot f\cdot (t-t_0)) \\
A_y\cdot\sin(2\pi\cdot f\cdot (t-t_0))  
\end{pmatrix}\cdot
\begin{pmatrix}
    \cos\theta & -\sin\theta \\
\sin\theta & \cos\theta
\end{pmatrix}+\begin{pmatrix}
    x_0\\y_0
\end{pmatrix}
\end{equation}
For each amplitude / frequency combination, we store the fit parameters and the statistical properties of the fit residuals. \autoref{appendix:residuals} shows the measured modulation time series and the residuals of the fit for different amplitudes and frequencies.

\subsection{Results}
\subsubsection{Amplitude of the modulation}
\begin{figure}
    \centering
    \includegraphics[width=1\linewidth]{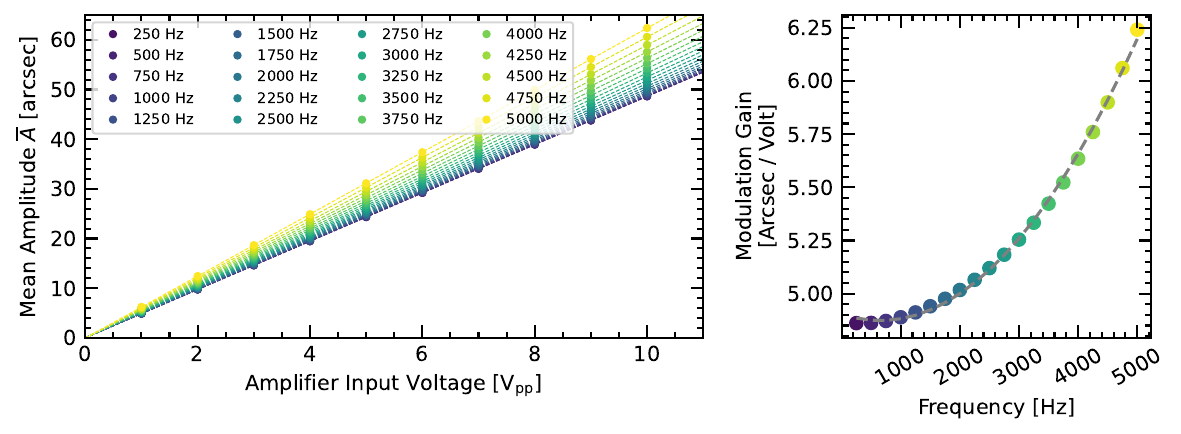}
    \caption{\textit{Left: } Measured modulation amplitude plotted against input voltage for the 20 different probed frequencies. \textit{Right:} Modulation gain plotted against the frequencies. The dashed line is a 2$^{\rm nd}$ order polynomial fit to the datapoints.}
    \label{fig:amplitude}
\end{figure}

From the fit of the measured modulation pattern (see previous section) we can derive several key parameters of the modulation pattern:

First, we determined the mean amplitude $\overline{A}=\frac{1}{2}(A_x+A_y)$ at the different 
input voltages and frequencies - see \autoref{fig:amplitude} (left). At all frequencies, the modulation amplitude shows a linear behaviour with the amplifier input voltage, demonstrating that the desired modulation amplitude can be easily controlled with the voltage. At the same time, the modulation $gain$, i.e. the slope of the relation between input voltage and modulation amplitude, shows an increase with frequency, that can be described by a quadratic function in the probed frequency range - see \autoref{fig:amplitude} (right).
\subsubsection{Ellipticity of the modulation}
\begin{figure}
    \centering
    \includegraphics[width=1.0
    \linewidth]{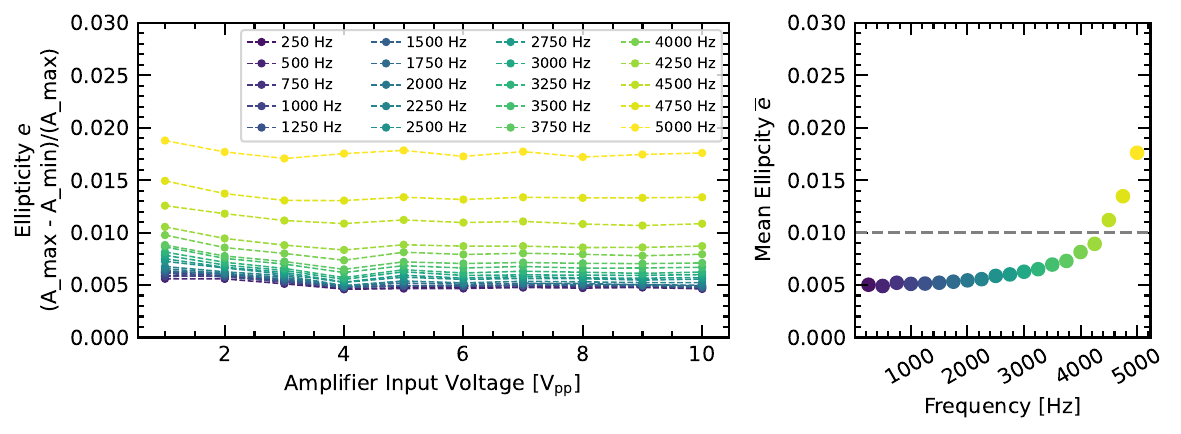}
    \caption{\textit{Left: } Ellipticity value $e$ plotted against amplifier input voltage for the 20 different probed frequencies. \textit{Right:} Mean ellipticity (averaged over all voltages) plotted against the frequencies.}
    \label{fig:ellipticity}
\end{figure}

Ideally, the modulation pattern is perfectly circular. In the next step of the analysis, we study the deviations of the measured modulation from the ideal circular pattern. \autoref{fig:ellipticity} shows how the ellipticity ($e=\frac{\max(A_x,Ay)-\min(A_x,A_y)}{\max(A_x,A_y)}$) changes with input voltage and frequency. Overall, the ellipticity shows no dependence on the amplifier input voltage and a monotonic increase with modulation frequency. For frequencies from 250\,Hz-4250\,Hz the ellipticity ranges between 0.005-0.010, after that it increases more steeply with frequency and reaches $e=0.018$ at 5\,kHz. 

\subsubsection{Non-elliptic deviations of the modulation}
\begin{figure}
    \centering
    \includegraphics[width=1\linewidth]{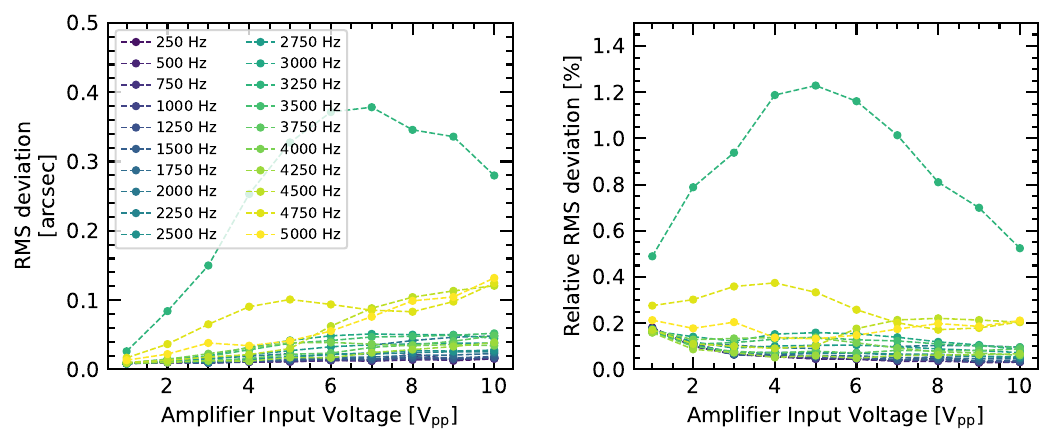}
    \caption{RMS of modulation residuals after the best-fit elliptical pattern has been subtracted. The \textit{left} plot shows the absolute values, while the \textit{right} plot shows the RMS relative to the respective modulation amplitudes.}
    \label{fig:rms}
\end{figure}

Finally, in \autoref{fig:rms} we study the fit-residuals of the modulation pattern. These residuals are a combination of both our measurement uncertainty and potential additional (non-elliptical) deviations from the commanded modulation pattern. For most commanded frequencies and amplitudes, the RMS of the fit residuals was below 0.05\,arcsec, or 0.2\% of the respective modulation amplitude, close to the precision of our individual measurements. However, there are notable exceptions: at high frequencies (4500\,Hz, 4750\,Hz, 5000\,Hz) the absolute RMS values reach up to 0.13\,arcsec and the relative RMS values reach around 0.3\%. In addition, the most extreme deviations can be seen at a frequency of $3250\,$Hz, where the absolute deviation reaches 0.38\,arcsec and the relative deviation reaches around 1.25\%. In \autoref{appendix:residuals} we show the measured residuals for several illustrative cases. We suspect that this behaviour at particular frequencies is caused by mechanical resonances, that lead to vibrations. 

\section{Summary and conclusions}\label{sec4}
\label{sec:summaryandconclusions}
Mechanical modulation of the light path is an essential ingredient for pyramid wavefront sensors and currently limited to modulation frequencies $\leq1$\,kHz. We have tested a novel modulator prototype capable of frequencies of up to 5\,kHz with a dedicated test setup. 
The modulator showed stable performance even at 5\,kHz with a maximum modulation amplitude of 62.4\,arcsec ($=302.5\,\mu$rad) with no measurable heat dissipation. 

For frequencies $<4.5\,$kHz the deviations from a perfectly circular modulation were minimal with a ellipticity $e<0.010$ and random deviations from the commanded modulation pattern RMS$<0.2\%$. At a single frequency (3250\,Hz) we measured larger residuals up to an RMS of 0.4\,arcsec/ 1.25\%, potentially related to mechanical resonances. For future revisions of the prototype it should be explored whether these resonances are internal to the modulation device or also related to its mounting.

 Potential future improvements of the modulator prototype could aim to reduce these resonances, by balancing moving parts and optimizing the housing. In addition, the potential vibrations that the modulator injects into the system should be quantified and, if neccesary, mitigated.

 For applications in real AO systems, the exact requirements on the mechanical performance, in particular on stability and repeatability of the modulation trajectory, depend on the respective instrument and the optical design involved. Existing instruments, like METIS and SAXO+, do require the trajectory to be repeatable to about 0.4 arcseconds (2 microrad).
This requirement is fulfilled by the modulator presented here in all cases, in fact even in the worst case at the 3250\,Hz resonance.

\section{Future work}
\label{sec:future}
We demonstrated stable operation and high modulation accuracy of the HF modulator prototype using a dedicated test setup. In future studies, the prototype could be integrated into AO test-benches to demonstrate its suitability for XAO applications on a system level. In AO simulations of potential future instruments, the influence of the measured deviations from ideal circular modulation should be taken into account and quantified.

It would also be important to quantify possible surface deformations of the modulator mirror at high frequencies.

In addition, the performance of the modulator prototype for other use-cases such as fast tip-tilt correction and non-circular modulation patterns could be explored.
\backmatter


\bmhead{Acknowledgements}
We thank the anonymous referee for constructive feedback that improved the presentation of our results.

The data reduction and visualization was performed using the following software packages: 

matplotlib \citep{2007CSE.....9...90H}, numpy \citep{2020Natur.585..357H}, scipy \citep{2020NatMe..17..261V}, IPhyton \citep{2007CSE.....9c..21P}, skikit-image \citep{scikit-image}

\section*{Declarations}
The modulator prototype was provided to MPIA by Physik Instrumente (PI) for testing. The authors do not declare any conflict of interest.








\begin{appendices}
\section{Plots of modulations measurements and their residuals}\label{secA1}
\label{appendix:residuals}
\begin{figure}
    \centering
    \includegraphics[width=0.9\linewidth]{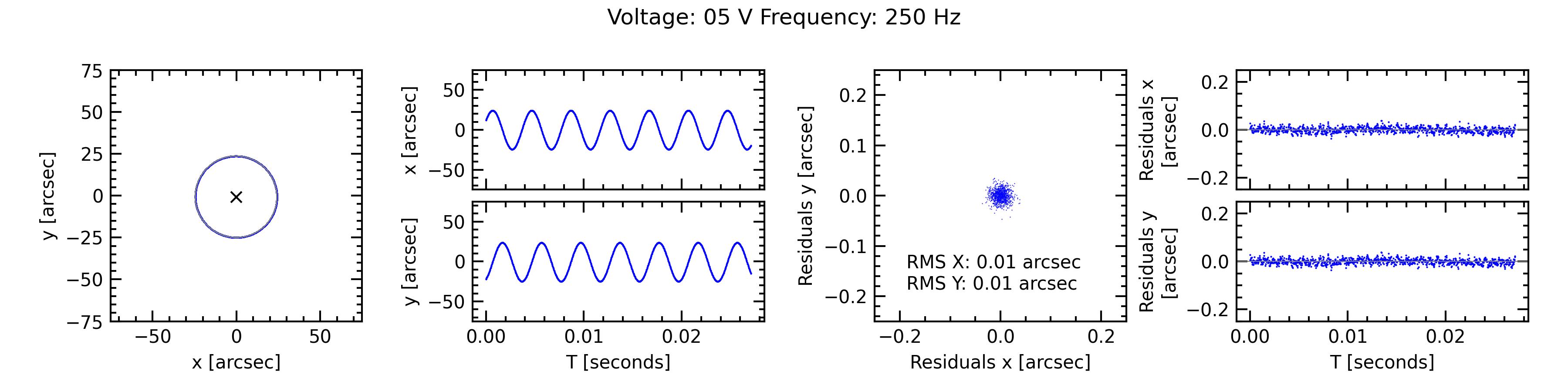}
    \includegraphics[width=0.9\linewidth]{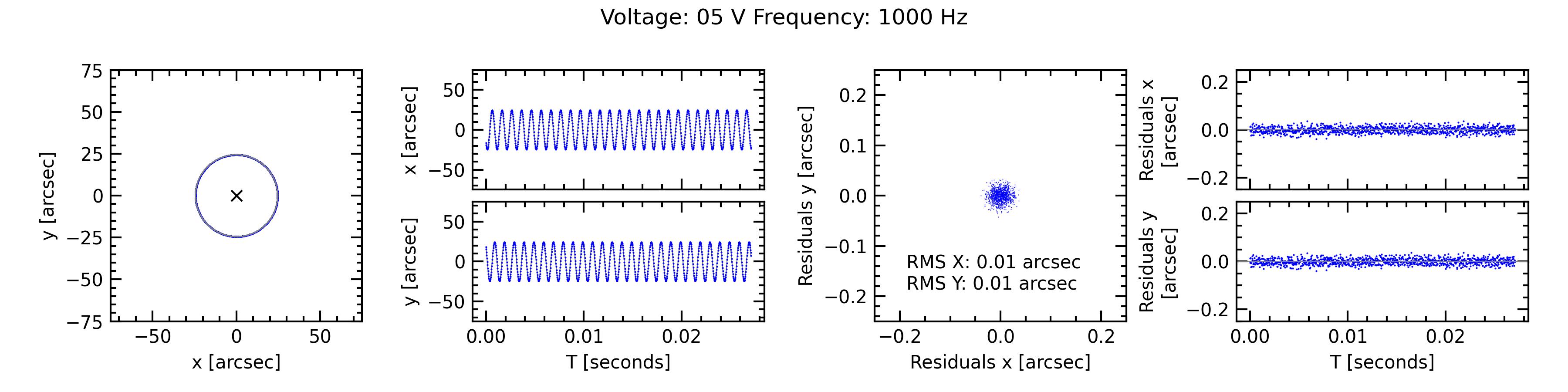}
    \includegraphics[width=0.9\linewidth]{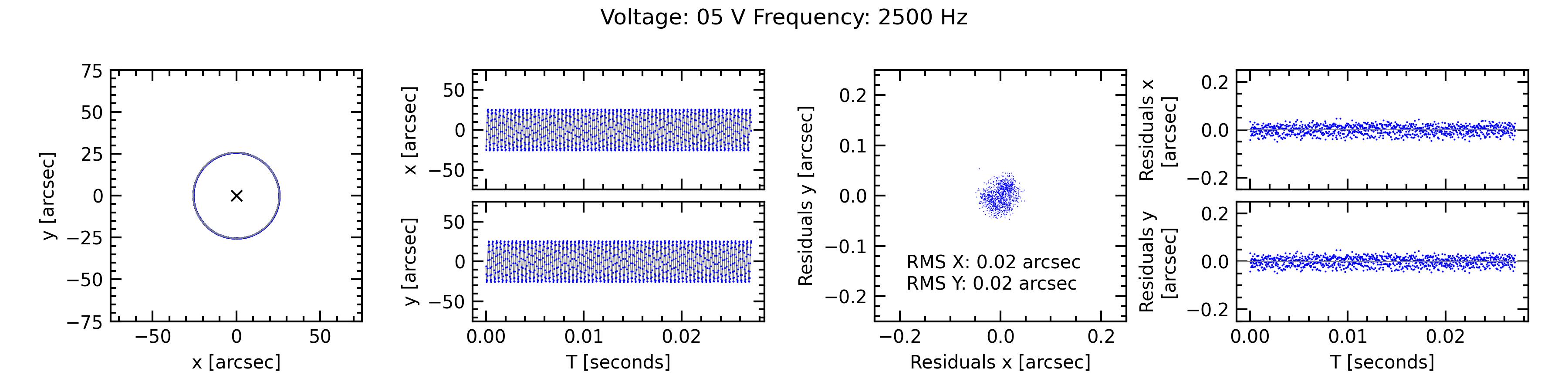}
    \includegraphics[width=0.9\linewidth]{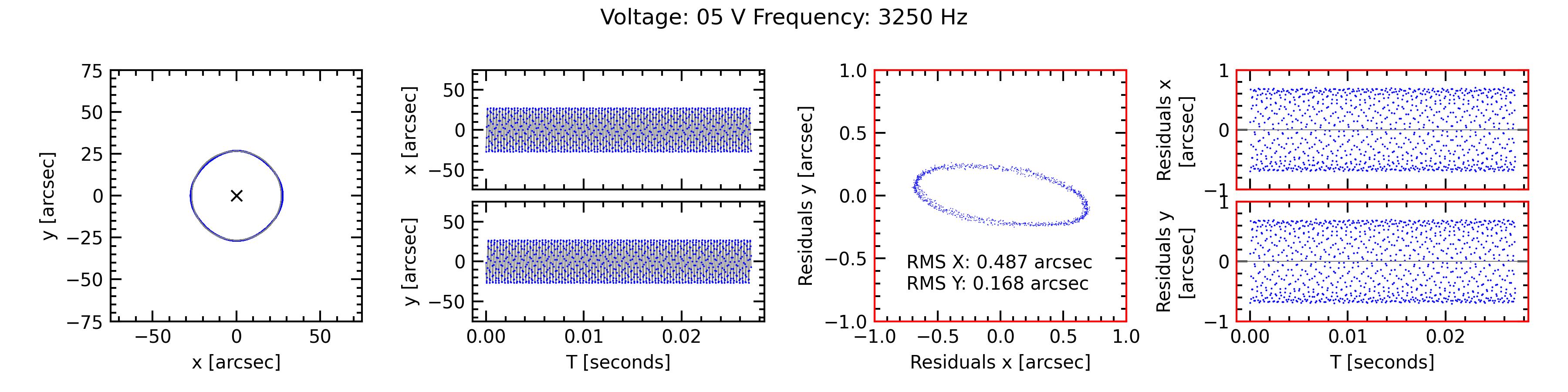}
    \includegraphics[width=0.9\linewidth]{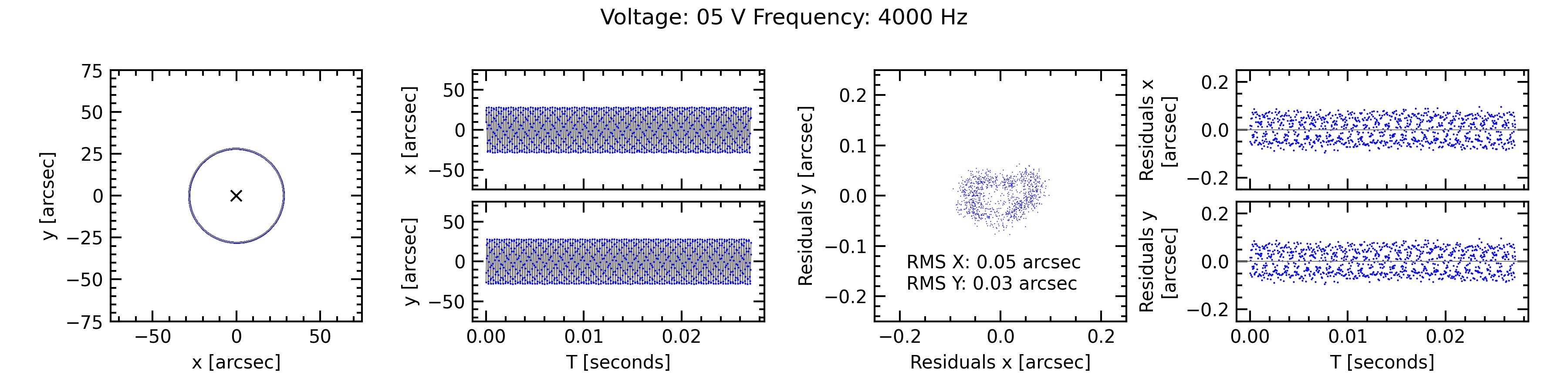}
    \includegraphics[width=0.9\linewidth]{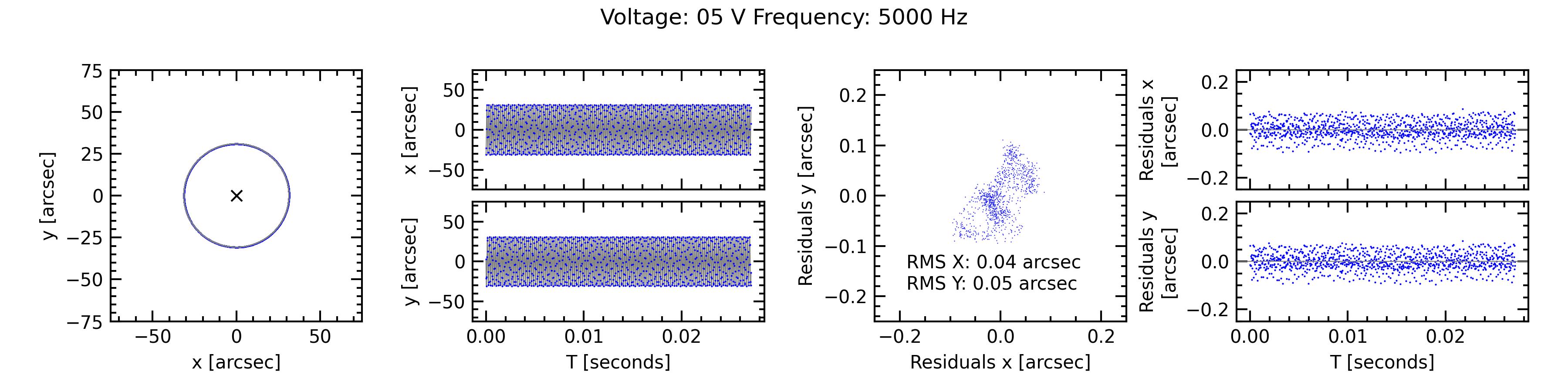}
    \caption{Measured modulation patterns (\textit{first and second column}) and residuals (\textit{third and fourth column}) from perfect elliptical modulation for six different frequencies and an amplifier input voltage of 5V$_{\rm pp}$, corresponding to about half of the maximum probed amplitude. Note the resonant behaviour at 3250 Hz. The different plot scale in the residuals is indicated by red axis spines.}
    \label{fig:residuals1}
\end{figure}

\begin{figure}
    \centering
    \includegraphics[width=0.9\linewidth]{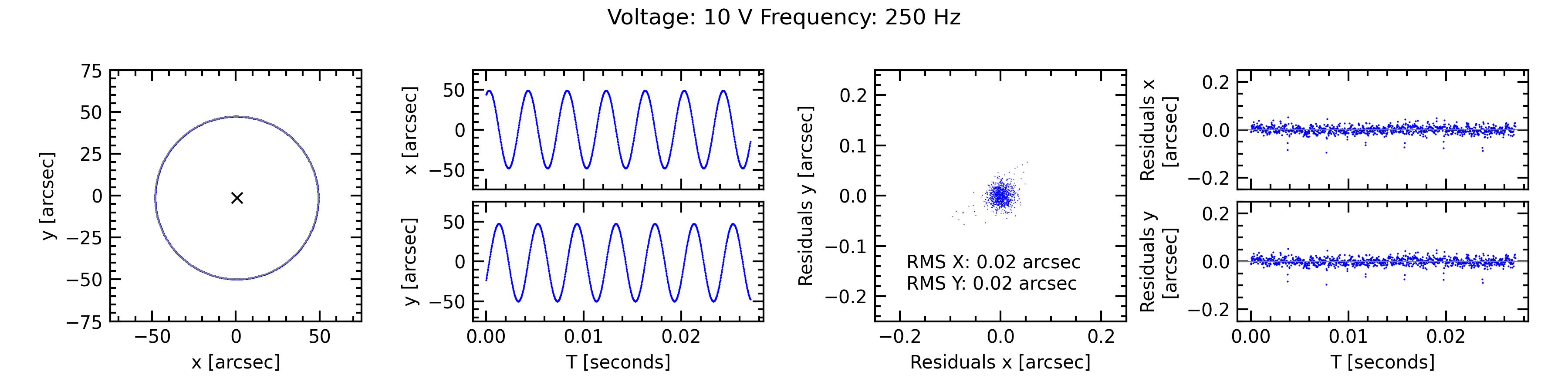}
    \includegraphics[width=0.9\linewidth]{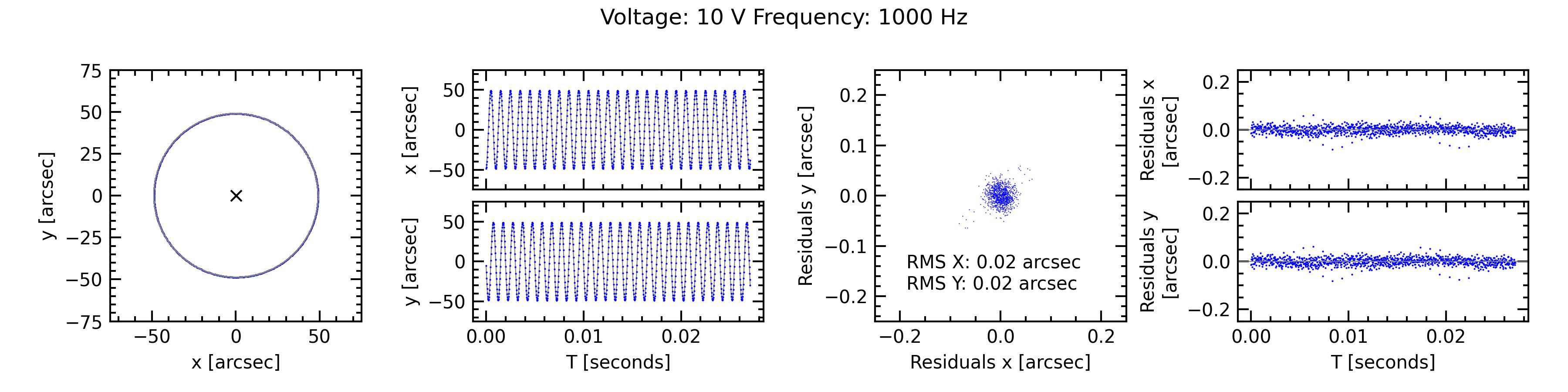}
    \includegraphics[width=0.9\linewidth]{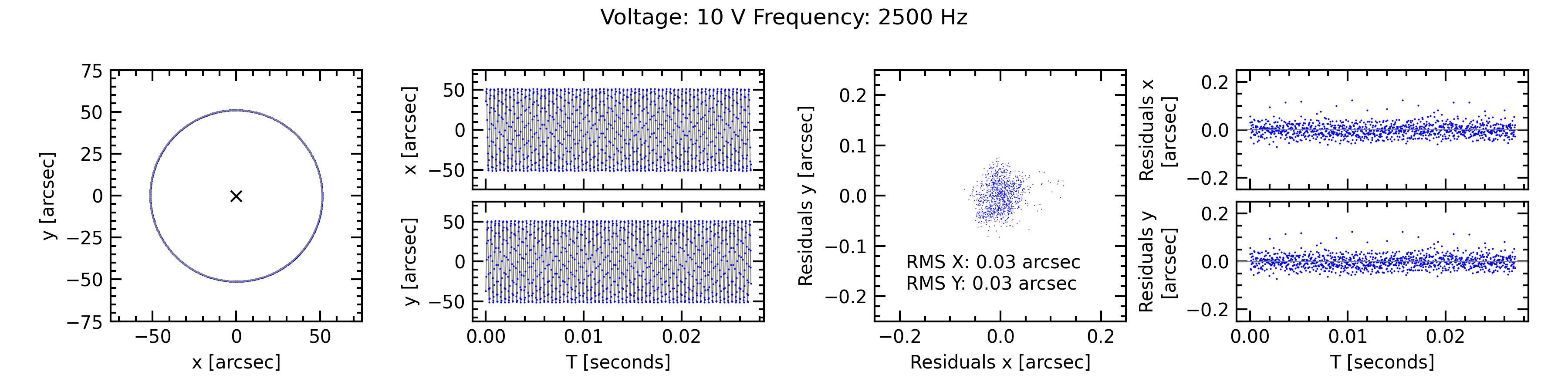}
    \includegraphics[width=0.9\linewidth]{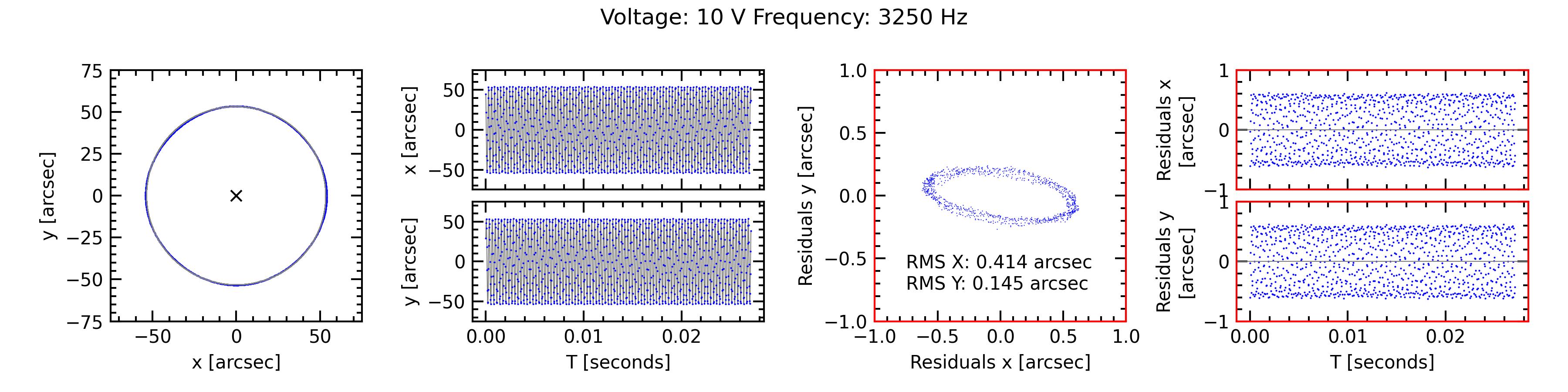}
    \includegraphics[width=0.9\linewidth]{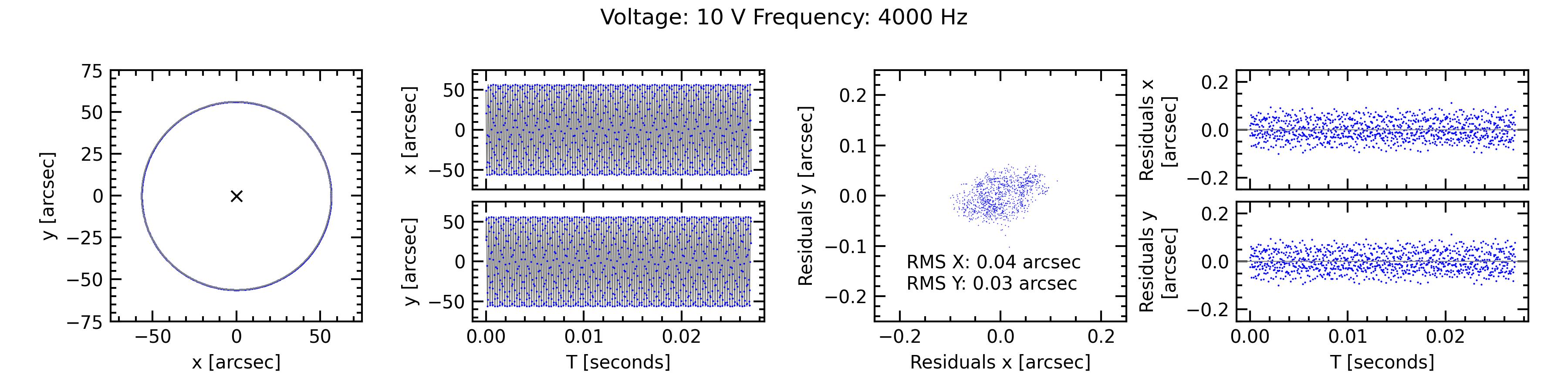}
    \includegraphics[width=0.9\linewidth]{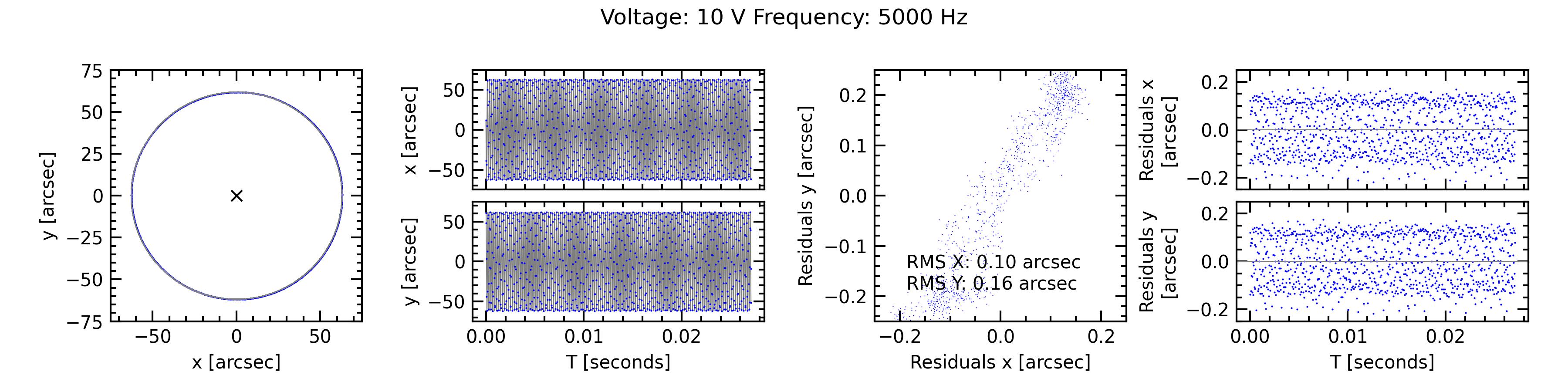}
    \caption{As \autoref{fig:residuals1}, but for the maximum probed amplifier input voltage (10V$_{\rm pp}$)}
    \label{fig:residuals2}
\end{figure}

The following figures show the measured modulator angles in $xy-$space (\textit{first column}) and as time-series (\textit{second column}) as well as the residuals of the ellipse fit (\textit{third and forth column}) for two amplitude settings (5V$_{\rm pp}$: \autoref{fig:residuals1}, 10V$_{\rm pp}$: \autoref{fig:residuals2}) and several selected modulation frequencies. We can make the following observations: At low modulation frequencies ($f\leq1000\,$Hz), the residuals look purely random and are close to the expected measurement noise (0.01 - 0.02 arcsec). The residuals start to increase for higher frequencies, and eventually reach a maximum at 3250\,Hz, which is likely a mechanical resonance and shows clear and systematic deviations from an elliptical pattern. At even higher frequencies, the residuals decrease again. We note that even though the systematic deviations from perfect modulation are clearly visible at high frequencies, their relative amplitude is small (even in the worst case at the resonant frequency it reaches just 1.2\%) and, therefore, they are unlikely to negatively effect the performance of a pyramid wavefront sensor.

\end{appendices}

\clearpage
\bibliography{sn-bibliography}


\end{document}